# IMPACT OF SOLENOID INDUCED RESIDUAL MAGNETIC FIELDS ON THE PROTOTYPE SSR1 CM PERFORMANCE*

D. Passarelli†, J. Bernardini, C. Boffo, S. Chandrasekaran, A. Hogberg, T. Khabiboulline, J. Ozelis, M. Parise, V. Roger, G. Romanov, A. Sukhanov, G. Wu, V. Yakovlev, Y. Xie,
FNAL, Batavia, IL 60510, USA

*Abstract*

A prototype cryomodule containing eight Single Spoke Resonators type-1 (SSR1) operating at 325 MHz and four superconducting focusing lenses was successfully assembled, cold tested, and accelerated beam in the framework of the PIP-II project at Fermilab. The impact of induced residual magnetic fields from the solenoids on performance of cavities is presented in this contribution. In addition, design optimizations for the production cryomodules as a result of this impact are highlighted.

## INTRODUCTION

The PIP-II project [1] has the scope to upgrade the existing Fermilab's accelerator complex to deliver the most intense high-energy beam of neutrinos for the international Deep Underground Neutrino Experiment (DUNE) at the Long Baseline Neutrino Facility (LBNF). It is based on a proton superconducting linac that comprises of five different Superconducting Radio Frequency (SRF) cavity types: 162.5 MHz half wave resonator (HWR), 325 MHz single spoke resonators type 1 and type 2 (SSR1, SSR2), low-beta and high-beta 650 MHz elliptical 5-cell cavities (LB650, HB650). International research institutions in India, the United Kingdom, Italy, France, and Poland are making significant contributions to the project, focused on accelerator technologies. Their contributions are playing a vital role in advancing and enhancing the project.

Positioned as the second cryomodule type in the linac, the two SSR1 cryomodules operate at a frequency of 325 MHz with continuous wave (CW) RF power and peak currents of 5 mA to accelerate H- beam from 10 MeV to 32 MeV. The PIP-II beam optics design requires that each SSR1 string assembly (see Fig. 1) contains four superconducting focusing lenses (solenoids) and eight identical SSR1 cavities [2, 3], where each cavity is equipped with one high-power RF coupler [4] and one frequency tuner [5].

The focusing lenses are combined superconducting magnets that include a solenoid with bucking coils and four corrector coils that can be power separately to provide both dipole and quadrupole steering. The focusing lenses are helium bath cooled and include resistive current leads optimized to minimize the heat load at 2 K. The design of the current leads feedthroughs was optimized to minimize potential frosting while avoiding the use of fans that could introduce microphonics issues. The geometry of the main and the bucking coils, the type and size of superconducting strand, and the number of turns in the winding packages were chosen to satisfy the minimum required integrated focusing strength of 4 $T^2 m$. The bucking coils are wound concentrically to the main coil and are located at each end of the lens. The maximum magnetic field generated by lenses in the cryomodule in the area near the surface of the SSR1 superconducting cavities is such that it will not result in more than two-fold reduction of the intrinsic quality factor after quench event at any point on its surface. However, no specification was defined for the maximum residual magnetic field induced of parts magnetized during the operation of the focusing lenses that would be trapped into the niobium surface during subsequent warmup and cooldown of the cryomodule. Fig. 2 illustrates the magnetic field distribution along the beam axis for the prototype SSR1 focusing lenses. It is evident that the fringe field extends significantly beyond the length of the focusing lens package, reaching magnitudes of thousands of Gauss outside the range of ±82.5 mm as depicted in the plot.

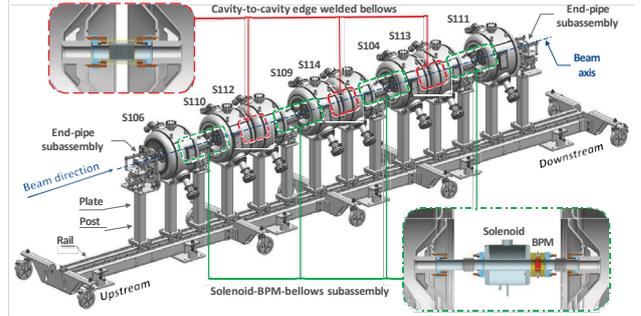

Figure 1: Layout of the string assembly for the prototype SSR1 cryomodule.

The prototype SSR1 cryomodule [6, 7] was designed to include a global, room temperature, magnetic shield positioned near the inner wall of the carbon steel vacuum vessel. The design and manufacturing of the global magnetic shield were aimed at reducing the impact of Earth's magnetic field to a level below 15 mG at the surfaces of the cavities. After the installation of the global magnetic shield in the vacuum vessel and the completion of the now standard demagnetization procedures [8], rigorous measurements validated the effectiveness of the shield's design to ensure that the desired performance requirement of magnetic field less than 15 mG along the beam axis (center of the vessel) was met.

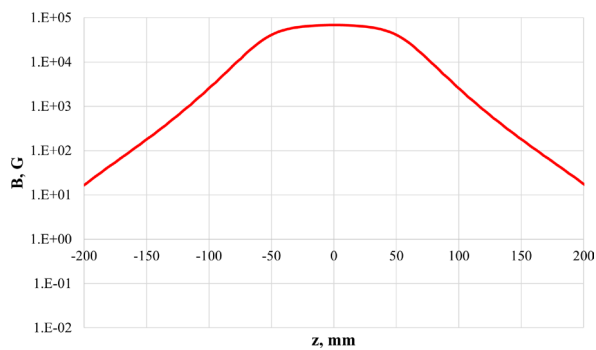

Figure 2: Magnetic field of the prototype SSR1 focusing lenses along the beam axis(z).

Prior to the assembly of the coldmass, all ferromagnetic parts and components located within the volume defined by the global magnetic shield were thoroughly measured to assess their residual magnetic fields. The measurement results demonstrated that these parts and components exhibited very low residual magnetic fields, measuring less than 5 mG at zero distance. In cases where any residual magnetization was detected, demagnetization procedures [8] were performed as necessary to minimize the magnetic fields to the desired levels. This rigorous process ensured that the components within the shielded volume had minimal magnetic interference and met the required specifications for the cryomodule assembly. Additionally, despite the high residual magnetic fields measured in the stainless steel helium vessels of the cavities, it was determined through cold testing in the spoke test cryostat that the performance of the cavities, particularly in terms of the quality factor ($Q_0$), exceeded specifications. As a result, it was decided to make an exception for the cavities and not pursue any further action to reduce the residual magnetic fields in the helium vessels.

The prototype SSR1 cryomodule (SSR1 pCM) was assembled and underwent cold testing at Fermilab between 2019 and 2021 [9]. During this period, measurements were conducted to assess the performance of the cryomodule in terms of accelerating gradients and dynamic heat load. This contribution compares test results of the cavities in the cryomodule with respect to tests of bare cavities in the vertical test stand and jacketed cavities in the spoke test cryostat (horizontal test stand).

## PERFORMANCE OF CAVITIES

Before string integration, a total of ten prototype SSR1 cavities were fabricated and tested to ensure the qualification of at least eight fully integrated jacketed cavities with high-power couplers and tuners. Initially, the cavities underwent cold testing as bare units in a vertical test stand to validate the manufacturing quality and chemical processing sequence. Subsequently, the cavities were jacketed with AISI 316L stainless steel helium vessels, processed, and subjected to cold testing in the spoke test cryostat for the final qualification with high-power couplers and tuners. The top-performing eight cavities were then selected for installation in the SSR1 pCM string assembly, see Fig. 1.

Finally, a phased approach was taken to cold test the cavities in the cryomodule. In Phase-I, the cryomodule was tested with the cavities operating at the required gradients for the first SSR1 cryomodule of the PIP-II linac. In Phase-II, the cavities' gradients were increased to meet the required values for the second SSR1 CM of PIP-II. Throughout both phases, the cavities underwent individual testing and full cryomodule unit testing to verify their performance.

Figure 3 provides a summary of the measured quality factor ($Q_0$) for the eight SSR1 cavities installed in the prototype module. The measurements were conducted in the following configurations:

- Vertical Test Stand (VTS): The cavities (with the exception of cavity S112) were tested and baselined as bare units in the VTS.

- Spoke Test Cryostat (STC): The cavities were tested as jacketed units with high power couplers and tuners in the STC.

- Cryomodule (CM): The cavities were tested in the cryomodule at the operating Phase-II gradients specified in Fig. 3.

To compare the $Q_0$ of cavities in the three configurations (VTS, STC, and CM), VTS and STC results taken at the specific accelerating gradient of 10 MV/m were used. This gradient value was selected as it closely matched the gradients used in Phase-II for the CM configuration (see bottom of Fig. 3). It is worth noting that all cavities were verified to be free from field emission during the measurements at 10 MV/m in both VTS and STC configurations. In the STC configuration, the administrative limit for the gradient was set at 14.4 MV/m, ensuring safe operation within the desired range. In the VTS configuration, the cavities were tested until quenching occurred, providing valuable information about their performance limits.

### $Q_0$ degradation from VTS to STC configuration

It is observed that there is a degradation in $Q_0$ when transitioning from the bare cavity configuration (VTS results) to the jacketed cavity configuration (STC results), with the exception of two cavities only (S110, S113). This degradation suggests that the presence of the helium vessel may introduce additional factors affecting the cavity performance. Measurements of the residual magnetic field at the surface of the 316L stainless steel helium vessels were conducted after the cold STC testing of each cavity. The purpose of these measurements was to assess the residual magnetic field resulting from the fabrication process of jacketing the cavity's niobium structure with 316L stainless steel helium vessels. Those measurements revealed significant residual magnetic fields up to 6 G in proximity to TIG welded joints between stainless steel components. These magnetic fields could explain the observed degradation in $Q_0$ from the vertical test stand (VTS) to the spoke test cryostat (STC) performance. However, despite the presence of residual magnetic fields, the average $Q_0$ value of the cavities was measured to be $9.9 \cdot 10^9$, exceeding the project's requirement of $Q_0 > 8.2 \cdot 10^9$. As a result, it was decided not to pursue any further action to attempt to demagnetize the helium vessels.

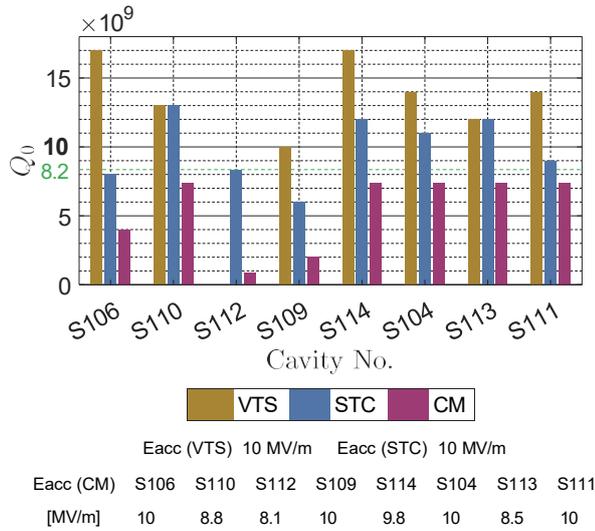

Figure 3: Comparison of the achieved $Q_0$ values for prototype SSR1 cavities in different configurations: Vertical Test Stand (VTS), Spoke Test Cryostat (STC), and Cryomodule (CM). The aim was to attain a minimum averaged $Q_0$ of 8.2 for cavities in the CM configuration at the specified CM gradients ($E_{acc}(CM)$).

## $Q_0$ degradation from STC to CM configuration

After undergoing cryomodule testing, it was observed that the performance of the cavities in the cryomodule configuration showed an additional degradation in $Q_0$ compared to their performance in the spoke test cryostat (STC) configuration. This degradation was attributed to several factors related to the cryomodule environment, with one significant factor being the presence of residual magnetic fields. Additionally, cavities S112 and S109 exhibited limitations due to radiation-induced field emission.

To investigate the impact of the cryomodule environment on the cavities, measurements were conducted to assess the residual magnetic fields at room temperature. These measurements were taken through the four tuner access ports positioned near cavities S112, S109, S113, and S111 (see Figs. 4), specifically in the vicinity of the second and fourth solenoids. The measurements were performed using A G93 Handheld 3-Axis Gaussmeter equipped with the Coliy G93 probe. The Gaussmeter had a measurement range of up to 20 T (200,000 G) and a resolution of 1 $\mu T$ (10 mG). To ensure precise readings, the probe was zeroed in the Zero Chamber before conducting the measurements. The purpose of these measurements was to evaluate the residual magnetic field induced by the solenoids when operated at cold temperatures and its effect on the surrounding ferromagnetic components.

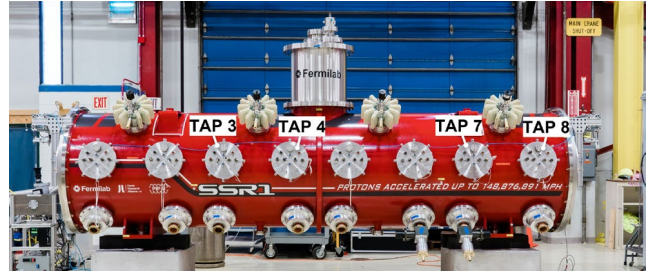

Figure 4: Highlighted are the Tuner Access Ports (TAP) used to access and perform measurement within the coldmass.

The results of these measurements, as shown in the Figs. 5, 6, revealed the presence of significant residual magnetic fields in the components surrounding the cavities. Notably, the magnetic field intensity was observed to increase near the solenoids, providing supporting evidence that the operation of solenoids at cold temperatures led to the permanent magnetization of ferromagnetic components within the cryomodule's cold mass. Additionally, during the cryomodule's warmup and cooldown cycle prior to Phase-II testing, it is likely that magnetic flux became trapped within the niobium material, contributing to the observed degradation in the cavities' quality factor ($Q_0$) when transitioning from the STC to the CM configuration.

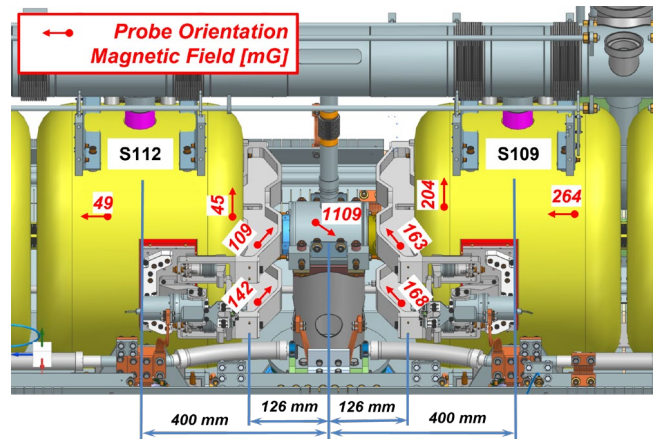

Figure 5: Measured residual magnetic fields in the components surrounding cavities S112 and S109.

To quantify the impact, the trapped magnetic field ($\Delta B$) was calculated for cavities without field emission in the SSR1 cryomodule. Cavities S112 and S109 were excluded from this study because affected by field emission. This calculation took into account the measured $Q_0$ values in both the STC and CM configurations, the G factor (84 W), and the sensitivity of surface resistance (Rs) to ambient magnetic field. The latter was obtained in STC with SSR1 cavity placed in the field of a pair of coils coaxial with the cavity axis and mounted at both ends around of helium jacket. The coils were energized with the cavity in normal conducting state, and then the cavity was cooled down to 2 K, allowing magnetic flux to be trapped in the cavity walls. Variations of Rs were measured with respect of the value of the magnetic field in the center of the cavity, and the Rs sensitivity was determined to be 0.05 $n\Omega/mG$. This was also confirmed by results presented for similar type of spoke cavities [10]. Table 1 provides the calculated values of the trapped magnetic field for the respective cavities.

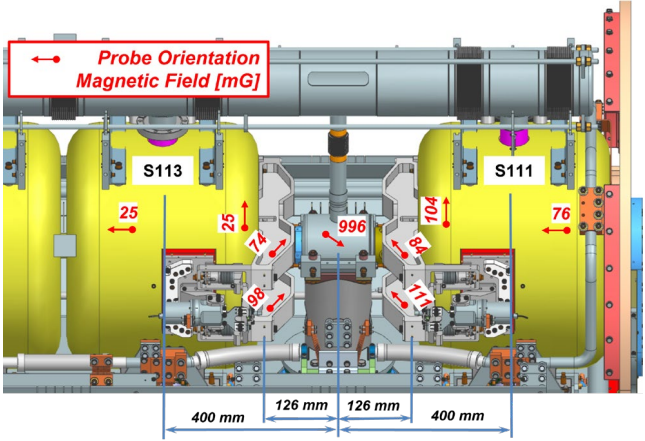

Figure 6: Measured residual magnetic fields in the components surrounding cavities S113 and S111.

It becomes evident that the cryomodule environment (residual magnetic fields) can significantly affects cavity performance. Mitigating these effects is crucial for optimizing the cavity's quality factor ($Q_0$) within the cryomodule.

Table 1

|  | S106 | S110 | S114 | S104 | S113 | S111 |
|---|---|---|---|---|---|---|
| $Q_0^{STC} \times 10^9$ | 8.00 | 13.0 | 12.0 | 11.0 | 12.0 | 9.00 |
| $R_S^{STC}\ [n\Omega]$ | 10.5 | 6.46 | 7.00 | 7.63 | 7.00 | 9.33 |
| $Q_0^{CM} \times 10^9$ | 4.00 | 7.40 | 7.40 | 7.40 | 7.40 | 7.40 |
| $R_S^{CM}\ [n\Omega]$ | 21.0 | 11.4 | 11.4 | 11.4 | 11.4 | 11.4 |
| $\Delta R_s\ [n\Omega]$ | 10.5 | 4.89 | 4.35 | 3.71 | 4.35 | 2.01 |
| $\Delta B\ [mG]$ | 210 | 97.8 | 87 | 74.3 | 87 | 40.4 |

## DEMAGNETIZATION ATTEMPTS

In an effort to mitigate the residual magnetic fields induced by the operation of the focusing lenses in the coldmass of the prototype SSR1 cryomodule, two demagnetization techniques were implemented. These techniques aimed to reduce the impact of the residual magnetic fields on the performance of the cavities and ultimately improve the quality factor ($Q_0$) of the cryomodule.

The first approach involved performing demagnetization cycles while operating the focusing lenses. The solenoid and correction currents were gradually ramped up to their maximum values (47 A and 30 A, respectively). Subsequently, they were ramped down in the opposite direction, with each subsequent ramp set at 75% of the absolute value of the previous step. This demagnetization procedure was based on successful techniques employed in other cryomodules, such as those used in the Facility for Rare Isotope Beams (FRIB) [11]. However, despite the application of these demagnetization cycles, no significant improvement in the $Q_0$ degradation was observed in the prototype SSR1 cryomodule. The degradation persisted both before and after the demagnetization process, suggesting that the residual magnetic fields were not effectively mitigated by this technique.

As the first technique did not yield the desired results, an alternative approach was pursued, involving the utilization of external coils on the vacuum vessel for demagnetization purposes. A power supply capable of delivering a maximum output of 65 A was employed to induce a peak field of approximately 1100 G inside the SSR1 vacuum vessel. The structure and location of the coils were optimized through the use of COMSOL simulations to maximize the effectiveness of the demagnetization process. The technique primarily focused on demagnetizing the global magnetic shield, as the fields generated were not sufficiently strong to effectively demagnetize components magnetized by the solenoids. However, similar to the previous technique, the improvements in the residual magnetic field of the cavities were marginal, and no detectable recovery of the $Q_0$ degradation was measured.

## CONCLUSIONS

The observed degradation in the quality factor ($Q_0$) of the prototype SSR1 cryomodule highlights the importance of thoroughly evaluating and addressing the effects of jacketing the cavities and the overall cryomodule environment even for low beta (0.22) spoke resonators operating at a low frequency of 325 MHz. This degradation in $Q_0$ is primarily caused by the remnant magnetic field at the surface of the SRF cavities, which leads to an increase in cavity residual resistance. The trapping of magnetic flux occurs as the cavity is cooled below its superconducting critical temperature (9.2 K). This trapped magnetic flux in the cavity walls results in increased residual resistance, leading to a degradation in $Q_0$.

It has been found that when an active source of magnetic field, such as a focusing lens, is situated near the SRF cavities within the coldmass, the implementation of a global magnetic shield and standard magnetic hygiene practices, which aim to and attenuate the Earth's magnetic field and to reduce residual magnetic fields before coldmass assembly, may not be sufficient to mitigate the detrimental effects on the performance of SRF cavities.

To optimize the performance of SRF cavities in the presence of high-fringe-field focusing lenses, lessons learned from the prototype SSR1 cryomodule and insights from the study presented in [12] have led to the integration of specific technical requirements into the design phase of production SSR1 and SSR2 cryomodules [13, 14] for PIP-II:

- Maximizing the use of non-magnetic materials like titanium, aluminum, and silicon bronze should be maximized. If the use of ferromagnetic materials cannot be avoided, such as in piping or tuners, the final parts shall have the lowest possible magnetic permeability (< 1.02). Materials such as 316L stainless steel or 316LN stainless steel (which shall be used as filler materials in welded joints as well) may meet this requirement, but it is important to consider provisions for annealing after welding and machining to restore the desired magnetic permeability.

- Setting a maximum residual magnetic field of less than 5 mG at zero distance when the focusing lenses are unpowered.

- Developing an ad-hoc demagnetization procedure using the focusing lens at cold operating conditions to effectively eliminate induced magnetization in the coldmass components. The objective is to achieve a maximum residual magnetic field below 5 mG before the subsequent warmup. If deemed necessary, the procedure shall be developed during the individual cold testing of the focusing lenses and prior to their integration into the sting assembly.

- Limiting the fringe field to below 10 G at the outer surfaces of an imaginary cylinder centered on the beamline axis, with a diameter of 0.70 m and length of 0.84 m, when the focusing lenses are powered.

- Considering the design of local shielding around the SRF cavities and/or the focusing lenses as an additional measure to mitigate the effects of focusing lens fringe fields.

## REFERENCES

[1] R. Stanek et al., *PIP-II Project Overview and Status*, Proceedings of SRF2023, Grand Rapids, Michigan, USA, 2023, MOIXA02.

[2] M.A. Hassan, et al., *Development of Low β Single-Spoke Resonators for the Front End of the Proton Improvement Plan-II at Fermilab*, IEEE Transactions on Nuclear Science (Volume: 64, Issue: 9, Sept. 2017).

[3] A. Sukhanov et al., *Characterization of SSR1 Cavities for PIP-II Linac*, Proceedings of SRF2019, Dresden, Germany, 2019, THP090.

[4] J. Helsper et al., *Design, Manufacturing, Assembly, and Lessons Learned of the Pre-Production 325 MHz Couplers for the PIP-II Project at Fermilab*, Proceedings of SRF2023, Grand Rapids, Michigan, USA, 2023, WEPWB094.

[5] D. Passarelli et al., *Performance of the Tuner Mechanism for SSR1 Resonators during Fully Integrated Testing at Fermilab*, Proceedings of SRF2015, Whistler, Canada, 2015, THPB061.

[6] T. Nicol et al., *SSR1 Cryomodule Design for PXIE*, Proceedings of IPAC2012, New Orleans, Louisiana, USA, 2012, WEPPD005.

[7] V. Roger et al., *Design update of the SSR1 cryomodule for PIP-II project*, Proceedings of IPAC18, Vancouver, Canada, 2018, WEPML019.

[8] S.K. Chandrasekaran, A.C. Crawford, *Demagnetization of a Complete Superconducting Radiofrequency Cryomodule: Theory and Practice*, IEEE Transactions on Applied Superconductivity, (Volume: 27, No. 1, 2017).

[9] D. Passarelli et al., *Test results of the prototype SSR1 cryomodule for PIP-II at Fermilab*, Proceedings of IPAC2021, Campinas, SP, Brazil, 2021, -THPAB343.

[10] D. Longuevergne, A. Miyazaki, *Impact of geometry on the magnetic flux trapping of superconducting accelerating cavities*, Physical Review Accelerators and Beams, 24(8), p.083101.

[11] K. Saito et al., *FRIB Solenoid Package in Cryomodule and Local Magnetic Shield*, Proceedings of SRF2019, Dresden, Germany, 2019, MOP072.

[12] R.E. Laxdal, *Review of Magnetic Shielding Designs of Low-Beta Cryomodules*, Proceedings of SRF2013, Paris, France, 2013, WEIOD01.

[13] J. Bernardini et al., *Final Design of the Production SSR1 Cryomodule for PIP-II Project at Fermilab*, Proceedings of SRF2023, Grand Rapids, Michigan, USA, 2023, WEPWB066.

[14] J. Bernardini et al., *Final Design of the Pre-Production SSR2 Cryomodule for PIP-II Project at Fermilab*, Proceedings of LINAC2022, Liverpool, UK, 2022, TUPOGE12.